\documentstyle[12pt,epsf,epsfig]{article}
\def\be{\begin{equation}}
\def\ee{\end{equation}}

\def\ba{\begin{eqnarray}}
\def\ea{\end{eqnarray}}
\def\ga{\mathrel{\raise.3ex\hbox{$$\kern-.75em\lower1ex\hbox{$\sim$}}}}
\def\la{\mathrel{\raise.3ex\hbox{$<$\kern-.75em\lower1ex\hbox{$\sim$}}}}

\begin{document}

\begin{titlepage}
\pagestyle{empty}
\baselineskip=21pt
\rightline{TPI--MINN--01/37}
\rightline{July 2001}
\vskip.25in
\begin{center}

{\large{\bf Search for Tracker Potentials in Quintessence Theory}}
\end{center}
\begin{center}
\vskip 0.4in

{Vinod B. Johri*}
\vskip 0.2in
{\it
{Theoretical Physics Institute, School of Physics and Astronomy, \\
University of Minnesota, Minneapolis, MN 55455, USA}}
\vskip 0.5in
{\bf Abstract}
\end{center}
\baselineskip=18pt \noindent
We report a significant finding in Quintessence theory that the the scalar fields 
with tracker potentials have a model-independent scaling behaviour in the expanding
universe. So far widely discussed exponential,power law or hyperbolic potentials can
simply mimic the tracking behaviour over a limited range of redshift. 
In the small redshift range where the variation of the tracking parameter
$\epsilon$ may be taken to be negligible, the  differential equation of generic potentials
leads to hyperbolic sine and hyperbolic cosine potentials which may approximate tracker 
field in the present day universe. We have plotted the variation of tracker  potential 
and the equation of state of the tracker field as function of the redshift $z$
for the model-independent relation derived from tracker field theory; we have also 
plotted the variation of $V(\Phi)$ in terms of the scalar field $\Phi$ 
for the chosen  hyperbolic cosine function and have compared with the curves obtained by 
reconstruction of $V(\phi)$ from the real observational data from the supernovae.
 .
\\[3mm]
\vspace*{18mm}
\begin{flushleft}
\begin{tabular}{l} \\ \hline
{\small Emails:vinodjohri@hotmail.com}
\end{tabular}
\end{flushleft}

\end{titlepage}
\baselineskip=18pt

The recent measurements \cite{1,2} of redshift and  luminosity distance relation of type 
Ia supernovae strongly suggest that the cosmic expansion is accelerating.
This provides an indirect evidence of the existence of an exotic matter , 
with positive energy density and negative pressure, dominant in the present day universe.
This exotic matter, termed quintessence, behaves like vacuum field energy and has repulsive 
(anti-gravitational) character due to its negative pressure.The dominance of the quintessence 
at the present epoch might very well account for the observed acceleration in the cosmic 
expansion. In fact, any physical phenomenon which generates negative pressure during cosmic 
evolution may give rise to quintessence. Some of the viable candidates for quintessence are 
as  follows:

i) scalar fields with time varying equation of state are the most favoured for quintessence.
Such fields acquire negative pressure during slow roll down of the scalar potential and have
been widely discussed \cite{3}-\cite{7} in the literature.

ii) Prigogine-type cosmological models \cite{8}-\cite{12} with matter creation out of gravitational 
energy. The universe is regarded as an open thermodynamical system wherein transformation of 
  gravitational energy into matter  induces negative pressure which supports expansion and might 
cause acceleration in cosmic expansion.

iii) cosmological models with dissipative phenomena where bulk viscosity comes into play.It is 
well known that the bulk viscosity can give rise to negative pressure in the standard 
FRW model \cite{13}-\cite{15}.
without affecting the isotropy of the universe. It is likely that the self-interaction 
of the cold dark matter in the galactic halos may cause bulk viscosity \cite{27} which might 
induce cosmic acceleration in the observable universe. Quintessence models with bulk viscosity 
have been considered by \cite {25,26}.

Long before the supernova observations, Ratra and Peebles \cite{19}, and Wetterich \cite {20}
had discussed the importance of the rolling scalar fields in the evolution of the universe.The
interest in scalar fields was revived with the luminosity distance - redshift observations of
type $I_a$ supernovae which suggest that about $70\%$ of the energy content of the universe 
consists of an exotic matter which induces acceleration in the cosmic expansion. Caldwell
et al \cite{3} discussed the possibilty that the scalar fields with evolving equation of 
state might constitute the exotic matter which counters gravitational attraction and supports 
expansion of the universe. But for the scalar energy density $\rho_\phi$ to be comparable with 
the matter energy density $\rho_n$ of the universe today, the initial conditions for the scalar 
fields must be set up carefully and fine-tuned. To get rid of the fine-tuning  and the coincidence 
problems, the notion of tracker fields \cite {16,17} was introduced. It permits the scalar fields
with a wide range of initial values of $\rho_\phi$ to rolldown along a common evolutionary track
 with $\rho_n$ to end up in the observable universe. But simply synchronized scaling of 
 $\rho_\phi$ with $\rho_n$ is not enough. A realistic 
 tracking behaviour must regulate the growth of $\Omega_\phi$ so that the additional 
 contribution of the scalar field to the energy density of the universe does not affect the 
 optimum Hubble expansion which yields the observed Helium abundance at nucleo-synthesis epoch. 
 Again it should not interfere with the process of the formation of galactic structure in the universe
 and must ensure transition from the matter to scalar field dominated era at the right epoch.
 To make quintessence physically viable, it should be linked to the cosmological observations. 
 With this objective, the notion of 'integrated tracking' was introduced \cite{18} which 
 essentially implies tracking compatible with astrophysical constraints. We shall briefly
 review the theory already discussed in \cite{18} to show that the tracker potentials 
 (independent of any particular choice) follow a definite path of evolution , in 
 compatibility with the observational costraints in the physical universe. The scientists
 have, so far proposed various scalar potentials with tracking properties, mostly in the 
 exponential, power law and hyperbolic forms but they represent the desired behaviour of tracking 
 over a limited range of redshift. In order to compare the model-independent tracking
 behaviour, predicted by our theory, with the observational  results based on SNe meaurements, 
 we assume a spatially flat universe with $\Omega_n = 0.3$ at the present epoch. It follows 
 from the tracker field theory that the transition to scalar field dominated era$(\Omega_n
 \simeq 0.5)$ corresponds to the value of tracking parameter $\epsilon = 0.666$ and it takes place at 
 $z=0.526$. We use interpolation techniques to calculate the values of $\Omega_\phi$ and 
 $\epsilon$ at different redshifts  during cosmic evolution and find that $\epsilon\simeq 0.98$
 at the present epoch

{\it Dynamics of Rolling Scalar Fields} -\,\,
Let us consider cold dark matter cosmology with quintessence in the form of 
 rolling  scalar fields,
with evolving equation of state, which acquire
repulsive character (owing to negative pressure) during the late evolution
of the universe. Such scalar fields would behave like $\Lambda_{eff}$ in the 
present day
observable universe,  and may turn out to be
the most
likely form of dark energy which induces acceleration in the cosmic
expansion.

Consider the homogeneous scalar field $\phi(t)$ which interacts with matter
only through gravity. The energy density $\rho_\phi$ and the pressure
$p_\phi$
of the field are given by
\be
\rho_\phi\, = \frac 12\,\dot\phi^2 + V(\phi)
\ee
\be
p_\phi\, = \frac 12\,\dot\phi^2 - V(\phi)
\ee
The equation of motion of the scalar field
\be
\ddot{\phi} + 3H\dot{\phi} + V'(\phi)\, = 0, \qquad
V'(\phi)\equiv\frac{dV}{d\phi}
\ee
leads to the energy conservation equation
\be
\dot{\rho}_\phi + 3H(1+w_\phi)\rho_\phi\, = 0
\ee
where $w_\phi\equiv\frac{p_\phi}{\rho_\phi}$ and
$H\equiv\frac{\dot{a}}{a}$
is
the
Hubble constant. Accordingly, $\rho_\phi$ scales down as
\be
\rho_\phi\,\sim {a}^{-3(1+w_\phi)} , \quad -1\leq w_\phi \,\leq1
\ee
Obviously, the scaling of $\rho_\phi$ gets slower as the potential energy
$V(\phi)$ starts dominating over the kinetic energy $\frac 12 \dot{\phi}^2$
of the scalar field and $w_{\phi}$ turns negative.

Since there is minimal coupling of the scalar field with matter/radiation,
it follows from Eq.(4) that the energy of matter /radiation is
conserved
separately as
\be
\dot\rho_n + 3H(1 + w_n)\rho_n = 0
\ee
Accordingly
\be
\rho_n\,\sim a^{-3(1+w_n)}
\ee
where $\rho_n$ is the energy density of the dominant constituent (matter or
radiation) in the universe with the equation of state $p_n = w_n\,\rho_n$
where $w_n = \frac 13$ for radiation and $w_n = 0$ for matter.

Although, the scalar field is non-interactive with matter, it affects the
dynamics of cosmic expansion through the Einstein field equations. Assuming
large scale spatial homogeneity and isotropy of the universe, the field
equations for a flat Friedmann model are
\be
 H^2 \, = \frac{\rho_n + \rho_\phi}{3m_p^2}
\ee
and
\be
 \frac{2\ddot {a}}{a}\, = -\frac{\rho_n+\rho_\phi +3p_n +3p_\phi}{3m_p^2}
\ee
where $m_p =2.4\times 10^{18}$ GeV is the reduced Planck mass.

Denoting the fractional density of the scalar field by $\Omega_\phi\equiv
\frac{\rho_\phi}{\rho_n+\rho_\phi}$ and that of the matter/radiation
field
by
$\Omega_n\equiv\frac{\rho_n}{\rho_n+\rho_\phi}$, equations (8) and (9)
may
be
rewritten as
\be
 \Omega_n + \Omega_\phi\, = \, 1
\ee
and
\be
2\frac{\ddot a}{a} = -\frac{\rho_n}{3m_p^2}\,[(1+3w_n)+(1+3w_\phi)
\frac{\Omega_\phi}{\Omega_n}]
\ee

The relative growth of $\Omega_\phi$  during the cosmic
evolution is given by
\be
\frac{\Omega_\phi}{\Omega_n} = \frac{\Omega_\phi^0}{\Omega_n^0}\,
\biggl(\frac{a}{a_0}\biggr)^{3\epsilon}
\ee
where the tracking parameter $\epsilon\equiv w_n - w_\phi$ and
$\Omega_\phi^0$, $\Omega_n^0$ denote the values of $\Omega_\phi$ and
$\Omega_n$
at the present epoch $ (a = a_0)$. As indicated by the recent supernovae
observations , $\Omega_\phi^0\simeq \frac{7}{3}\Omega_n^0$ at the present epoch; 
consequently Eq.(12) may be expressed in terms of the
red-shift $z$ as below
\be
(\Omega_\phi^{-1} - 1) =\, 0.43(1+z)^{3\epsilon}
\ee

If we insist that the scalar field, regardless of its initial value, should
behave like $\Lambda_{eff}$ today, it must obey tracking conditions
\cite{16,17,18} which have wide
 ramifications for quintessence fields as already discussed in detail
\cite{16,18}. In nutshell, tracking
consists in synchronised scaling of $\rho_\phi$ and $\rho_n$ along a common
evolutionary track with $w_\phi< w_n$ so as to ensure the restricted growth of $\Omega_\phi$
during the cosmic evolution in accordance with the observational  constraints.

 {\it Search for Tracker Potential} - \,\,
We have already discussed \cite{18} the theory of integrated tracking in brief 
and shown that the scalar fields can give rise to quintessence ($w_\phi < 0$ )
during the slow rolldown of scalar potential when the kinetic energy term is 
very small compared to $V(\phi)$.
Then from Eq.(1) and (2)
\be
      \frac{\dot\phi^2}{\rho_\phi}\, = 1+w_\phi 
\ee
and
\be
      V(\phi) \, = \Bigl(\frac{1-w_\phi}{2}\Bigr)\rho_\phi\,\sim\,\rho_\phi
\ee

Thus $V(\phi)$ scales down effectively as $\rho_\phi$ during tracking.
Differentiating Eq.(14) logarithmically with respect to time, we get the slow 
roll down condition for the scalar potential under the assumption  that $\dot{w_\phi}\simeq0$

\be
\pm\frac{V'(\phi)}{V(\phi)} =\frac{3H(1+w_\phi)}{\dot\phi}=\sqrt\frac{3(1+w_\phi)}{\Omega_\phi m_p^2}
\ee

Again differentiating logarithmically and writing $\zeta$ for $\frac{V'}{V}$,we get

\be
      \mp \frac{3H(1+w_\phi) \zeta'}{\zeta^2}\, =\,\frac{\dot\Omega_\phi}{2\Omega_\phi}
\ee

Logaithmic differentiation of Eq.(12) with respect to time yields

\be
\frac{\dot\Omega_\phi}{\Omega_\phi(1-\Omega_\phi)}\,= 3(\dot\epsilon ln a + \epsilon H)
\ee

Combining Eq.(17) with Eq. (18) and inserting the value of $\Omega_\phi$ from Eq.(16),
we get

\be
\mp\frac{\zeta'}{\zeta^2 - k^2}\,=\,\frac{\dot\epsilon ln a +\epsilon H}{2H(1+w_\phi)}      
\ee

Using the criterion $\dot\epsilon\geq0$ for tracking \cite{18}, the generic tracker 
potentials are given by the differential equation

\be
       \mp\frac{\zeta'}{\zeta^2 - k^2}\,\geq \, \frac{\epsilon}{2(1+w_n -\epsilon)}
\ee

where $+\zeta'$ corresponds to decreasing potentials and $-\zeta'$ to increasing potentials
and $k^2 = 3(1+w_\phi)/{m_p}^2$
According to our notation, the prime denotes differentiation with respect to $\phi$ and an 
overdot denotes time-derivative.

The dynamics of tracking depends sensitively on the variation of the tracking parameter
$\epsilon$ during cosmic evolution. For this reason, integrated tracking links $\epsilon$
to the observational constraints \cite{18}. For example, the choice of $\epsilon = 0.666$ at
$z= 0.526$ corresponds to $\Omega_\phi \simeq 0.5$ which ensures the onset of acceleration 
in cosmic expansion 
 at this  epoch. Again, $\epsilon\leq 0.035$ at $z= 10^{10}$ corresponds to $\Omega_\phi<0.14$
 which ensures that the observed 
helium abundance by nuleosynthesis, as successfully predicted by the standard model, is
not disturbed by the presence of the scalar field. 
 It is noteworthy that the value of the redshift marking the onset of acceleration depends 
 upon the observed
 value of $\Omega_n^0$. The transition to scalar dominated era will occur at $z= 0.414$ for 
 $\Omega_n^0 = 0.35$ and at $z= 0.732$ for $\Omega_n^0 = 0.25$\,
 By using interpolation techniques \cite{21}, 
the complete tracking profile of the scalar fields can be mapped for any range of redshift. 
In particular for the small redshift range, where the variation in the tracking parameter 
$\epsilon$ is negligible, 
 two exact solutions 
of Eq.(20) may be obtained as follows:

Case I.  Tracker Field with Decreasing Potential

Taking +ve sign with $\zeta'$, Eq,(20) gives on integration,

\be
    \frac{V'}{V}\equiv\,\zeta\,=-k\,{coth\, k(\frac{\phi}{\alpha} +\beta)}
\ee
which yields on further integration

\be
         V \, = \,A{sinh \,k(\frac{\phi}{\alpha} + \beta)}^{-\alpha}
\ee
where $\alpha = \frac{\epsilon}{2(1-\epsilon)}$ and $w_n=0$ in the small redshift range under 
consideration.

This is the analytical derivation of the scalar potential proposed by Urena-Lopez and Matos
\cite{22} for tracker fields but it holds good for constant tracking parameter.

Case II. Tracker Field with Increasing Potential

Taking -ve sign with $\zeta'$, Eq.(20) gives on integration,

\be
    \frac{V'}{V} \equiv\,\zeta\, = k\, [ tanh\, k(\frac{\phi}{\alpha} -\beta)]
\ee
which yields the integral solution

\be
     V\, =\, A[cosh \,k(\frac{\phi}{\alpha} - \beta)]^{\alpha}
\ee

The tracker potential (24) is of the same form as proposed by Sahni and Wang \cite{23}.
We shall be using it for comparison with the reconstructed potential from the observational 
data\, \cite{24}.
\begin{figure}
\begin{center}
\mbox{\epsfig{file=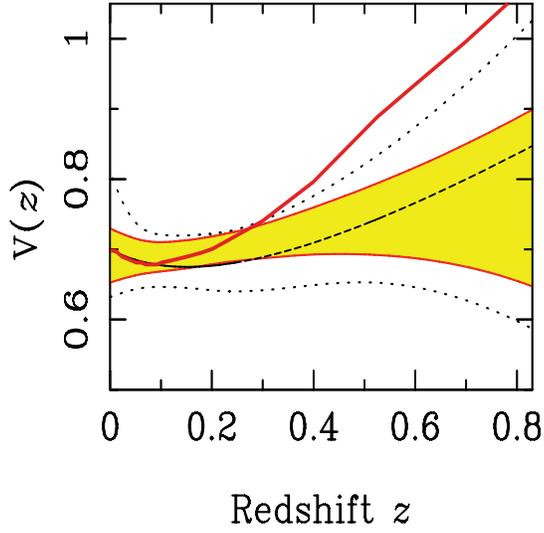,height=8cm,width=8cm}}
\end{center}
\caption[.]{\label{fig:figj}\it
The tracker potential  $V(z)$, shown in units of the critical density
$\rho_{cr}^0$, is plotted as a function of the redshift. The solid line shows the potential from tracker theory and the thin blue line shows the potential reconstructed from the observational data}
\end{figure}
\begin{figure}
\begin{center}
\mbox{\epsfig{file=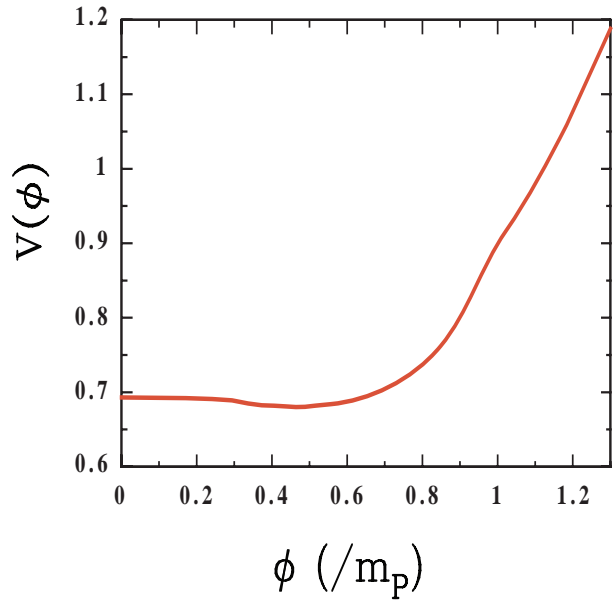,height=8cm,width=8cm}}
\end{center}
\caption[.]{\label{fig:figj3}\it
The tracker potential $V(\Phi)$ is shown in units of the critical density $\rho_{cr}^0$ at the present epoch. The value of $\Phi$ is shown in units of planck mass.}
\end{figure}
\begin{figure}
\begin{center}
\mbox{\epsfig{file=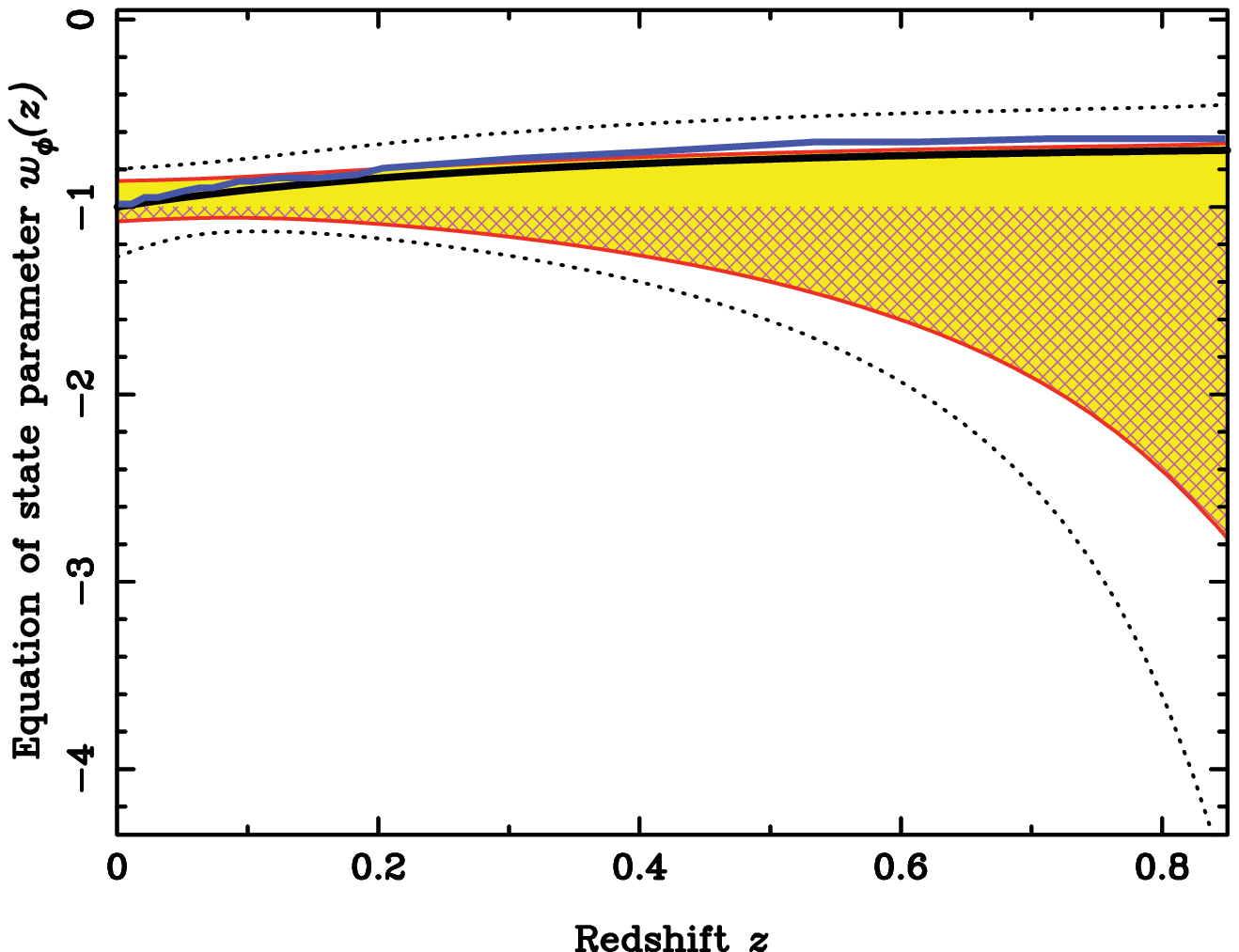,height=8cm,width=8cm}}
\end{center}
\caption[.]{\label{fig:figj2}\it
The equation of state parameter $w_\phi(z)=\frac{p_\phi}{\rho_\phi}$ as a function of the redshift $z$. The solid blue line depicts the variation according to tracker field theory, the black line shows the variation corresponding to the reconstructed equation of state from the observational data.}
\end{figure}

{\it Comparison with Observational Data} -\,\,
First we use Eq.(15) to show that the tracker potential $V(\phi)$  
may be expressed as a function of redshift, in conjunction with Eqs.(5),(7) and (13),as

\be
    V(z) \, = \, \bigl(\frac{1+\epsilon -w_n}{2}\bigr)(1+z)^{3(1-\epsilon)}\rho_\phi^0 
\ee

in the model independent form where $\epsilon$ conforms to astrophysical constraints and its 
value at different redshifts is known to us by interpolation \cite{21}. Accordingly, the
$V(z)$ curve plotted from Eq.(25) is the characteristic curve for all tracker potentials.
It leads to a significant result in Quintessence theory that 
{\bf the tracker potentials scale 
down along a definite path in the expanding universe independent of the functional 
form of $V(\phi)$}.

Throughout the matter dominated and  scalar field dominated era, 
$w_n=0$ and $\epsilon=\, -w_\phi$.
With this simplification, we can plot the $V(z)$ curve by writing Eq.(25) in the form

\be
\frac{V(z)}{\rho_{cr}^0 }\, = 0.35(1+\epsilon)(1+z)^{3(1-\epsilon)}
\ee

where $\rho_{cr}^0$ is the critical density of the universe at the present epoch.
The plotted curve is shown in figure 1.

Let us now examine as to how far the the potential function (24) agrees with the form of 
$V(\phi)$ reconstructed by Saini et al  \cite{24} from the real observational data of the 
supernovae \cite{1,2}. We can express potential in units of the critical density and $\Phi$ 
 in units of planck mass and rewrite Eq. (24) in the form

\be
   \frac{V(\Phi)}{\rho_{cr}^0} \, = \, 0.35(1+\epsilon)\Bigl[\cosh\frac{\Phi}{m_p}\Bigr]^\alpha
\ee

taking $\Phi=0$ when $z=0$ (since in case of increasing potential ($\frac{dz}{d\Phi}>0$) where
$\frac{\Phi}{m_p}\equiv k(\frac{\phi}{\alpha} -\beta)$. The $V(\Phi)$ curve(figure 2) has been  be plotted  from Eq.(27)
by calculating $\frac{\Phi}{m_p}$ in terms of the redshift from the relation

\be
    \cosh(\frac{\Phi}{m_p}) = (1+z)^{3\epsilon/2}
\ee
The $w_\phi(z)$ curve (fig.3) is plotted from the interpolated values of $\epsilon$ for different
values of $z$.

 Although the variation of $w_\phi$ as a function of the redshift $z$, predicted from the tracker
  field theory, lies within 
 $1\sigma$ range when compared with
the corrersponding curve obtained by Saini et al\, \cite{24}, the variation of the tracker 
potential V as a function of $z$ and as a function of $\phi$
 appears to have some deviation in the rolldown behaviour. We should 
like to point out that the reconstruction of $V(\phi)$ from the observational data by
Saini et al \cite{24} and our theoretical results derived from tracker field theory both are
based on the assumption  $\Omega_n^0 = 0.3$. This presumption leads to the interpolated 
value $w_\phi = -0.98$ at the 
present epoch.It implies that the present value of the scalar field energy density is very close to
the cosmological constant $\Lambda$ with ($w_\phi = -1$) and the future expansion of the 
universe will be given
by the scale factor $a\sim \sinh^\frac{2}{3}(\frac{3}{2}\sqrt{\Lambda/3}ct)$    
However, the initial choice of $\Omega_n^0 = 0.33$  leads to $w_\phi = -0.77$ at the present 
epoch and a different tracking path for the tracker potential but it remains model-independent
as given by Eq.(25)    . The matching of our model with the reconstructed $V(\phi)$ lends 
support to 
the view that the scalars fields might be the ultimate choice for the quintessence energy.

{\large\bf Acknowledgments}
This work was supported  by UGC grant from India. The author
acknowledges useful
discussions and valuable help of Keith Olive, Panagiota Kanti and Tonnis Ter Veldhuis and
hospitality of
Theoretical Physics Institute, University of Minnesota.

* Permanent Address: Department of Mathematics and Astronomy, Lucknow
University,
 Lucknow 226007. India.

\end{document}